\documentclass[%
preprint, %
]{revtex4}

\usepackage{bm,amssymb,amsmath}
\usepackage{graphicx}
\usepackage{wrapfig}

\newcommand{\p}{\partial}
\newcommand{\EB}{\bm{E} \times \bm{B}}
\newcommand{\imag}{\mathrm{i}}
\newcommand{\diff}{\mathrm{d}}
\newcommand{\odf}[2]{\frac{\diff #1}{\diff #2}}
\newcommand{\pdf}[2]{\frac{\partial #1}{\partial #2}}
\newcommand{\odft}[2]{\frac{\diff^{2} #1}{\diff #2^{2}}}

\begin{document}

 \title{Bifurcation in electrostatic resistive drift wave turbulence}
 \author{Ryusuke Numata}
 \email{ryusuke.numata@anu.edu.au}
 \author{Rowena Ball}
 \author{Robert L. Dewar}
 \affiliation{Department of Theoretical Physics, Research School of
 Physical Sciences and Engineering, The Australian National University,
 Canberra, ACT 0200, Australia}

 \begin{abstract}
  The Hasegawa--Wakatani equations, coupling plasma density and
  electrostatic potential through an approximation to the physics of
  parallel electron motions, are a simple model that describes resistive
  drift wave turbulence.  We present numerical analyses of  bifurcation
  phenomena in the model that provide new insights into the interactions
  between turbulence and zonal flows in the tokamak plasma edge region. 
  The simulation results show a regime where, after an
  initial transient, drift wave turbulence is suppressed through zonal flow
  generation.  As a parameter controlling the strength of the
  turbulence is tuned, this zonal flow dominated state is rapidly
  destroyed and a turbulence-dominated state re-emerges.  The transition
  is explained in terms of the Kelvin-Helmholtz stability of zonal
  flows.  This is the first observation of an upshift of turbulence
  onset in the resistive drift wave system, which is analogous to the
  well-known  Dimits shift in turbulence driven by ion temperature gradients.
   \end{abstract}

\maketitle
\section{Introduction}
\label{sec:introduction}

Fusion plasmas and other turbulent flows in quasi-two-dimensional (2D) 
geometry can undergo spontaneous transitions to a turbulence-suppressed
regime. In plasmas they are known as \emph{L--H} (low-to-high
confinement) transitions and are studied intensively because they
effectively enhance the confinement, through suppression of anomalous
or turbulent particle and heat fluxes. It is now widely accepted that 
emergent zonal flows are crucial to achieving confinement improvement
\cite{Diamond-Itoh-Itoh-Hahm_05}. The \emph{L--H} transition is
associated with nonlinearly self-generated poloidal $\EB$ shear or
zonal flows \cite{transition_by_external_flow} in the tokamak edge
region, which comprises the transition zone from inner hot core plasma
to the outer cold scrape-off layer. Zonal flows reduce anomalous
transport by absorbing energy from drift waves and by shearing apart
eddies which mediate turbulent transport, and thus play a key role in
its regulation.

In this paper we present the results of analytic and numerical
investigations of transitions between turbulence-dominated and
zonal-flow-dominated regimes, using the Hasegawa--Wakatani (HW) model
\cite{Hasegawa:1983,Wakatani_Hasegawa_84} for electrostatic resistive
drift wave turbulence in 2D slab geometry. We find that bifurcations in
the model correspond to the onset of drift wave turbulence, the
generation of zonal flows, and the re-emergence of turbulence as the
zonal flows become unstable, and observe that this is drift wave
turbulence analog of the Dimits shift \cite{Dimits_etal_00} in ion
temperature gradient turbulence.
 
Three energetic subsystems interact to produce the complexity observed
in \emph{L--H} transition dynamics: the kinetic energy of turbulence,
the kinetic energy of shear flows, and the potential energy contained in
density or pressure gradients. The three major governing processes are
generation of turbulence by drift waves, self-organization of zonal
flows, and destabilization of the zonal flows. The instabilities that
lead to these changes correspond to bifurcations of equilibrium 
solutions of model equations.  If a tunable parameter crosses a
stability threshold the qualitative nature of the solution changes. We
say that a \emph{primary instability} occurs at a linear stability
threshold of the equilibrium with zero background flow, which physically
corresponds to the onset and growth of drift waves. Theoretical
\cite{Biskamp:1985} and experimental \cite{Klinger:1997} studies have
indicated that the generation of drift wave turbulence in plasmas may
occur by the Ruelle-Takens mechanism \cite{Ruelle:1971}, in which a
limit cycle generated by a Hopf bifurcation undergoes a Niemark-Sacker
bifurcation to a torus, which may undergo one or more bifurcations to
higher-dimensional tori before the motion becomes chaotic. 

However, to complicate this generic turbulence onset scenario, in
plasmas zonal flows will be generated beyond the primary threshold due
to an instability of the drift waves, effectively suppressing drift wave
activity. This instability causing the zonal flow onset is termed a
\emph{secondary instability}. We can consider the turbulence to be
well-developed at the secondary instability; i.e., for heuristic
purposes we assume the Ruelle-Takens sequence to have already occurred.

A strong candidate for this secondary instability mechanism is
modulational instability
\cite{Guzdar_Kleva_Chen_01,Dewar_Abdullatif_07}, a special case of
\emph{nonlinear} mode coupling whereby modulation of a small scale
monochromatic wave can transfer energy non-locally to a longer
wavelength structure due to the ponderomotive force effect leading to
excitation of zonal flows. One might also expect an inverse energy
cascade, endemic to quasi two-dimensional flows in general, whereby
local mode coupling channels energy into large scale structures. 

A different mechanism for this secondary instability that generates
zonal flows is Kelvin--Helmholtz (KH) instability
\cite{Rogers_Dorland_Kotsch_00,Jenko_Dorland_Kotsch_Rogers_00}. 
In this scenario the KH instability may be driven by radially elongated
drift wave eigenmodes. The KH mode of the drift waves necessarily
possess a zonal flow component, and provide a natural mechanism for the
zonal flow growth. 

\begin{figure}
 \centerline{\includegraphics[scale=1]{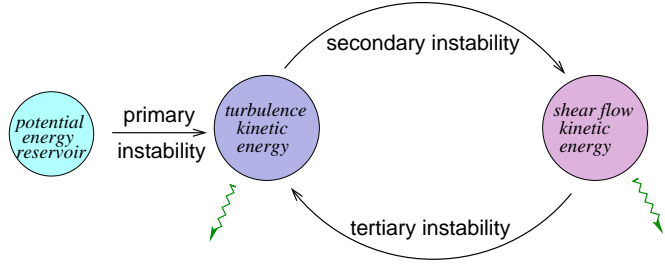}}
 \caption{\label{energy-flux} Primary instabilities generate turbulence
 from a potential energy reservoir, secondary instabilities lead to the
 growth of shear or zonal flows at the expense of turbulence kinetic
 energy, and tertiary instabilities may destabilize the shear or zonal
 flows. Zigzag green arrows represent dissipative channels.}
\end{figure}

As the zonal flows become more energetic they are subject to
\emph{tertiary instability} which breaks up the coherent zonal
structuring of the flow into turbulent small scale eddies via KH
instabilities of the zonal flows. The small scale turbulence may again
cohere via secondary instabilities. These interactions are schematized
in Fig.~\ref{energy-flux}.

Nonlinear interactions between zonal flows and drift waves results in an
upshift of the boundary in parameter space for the tertiary onset of
turbulence. This is known as the Dimits shift in
ion-temperature-gradient (ITG) driven turbulence, and the turbulence
suppressed regime was mapped by gyrokinetic and gyrofluid simulations
\cite{Dimits_etal_00}. 

The simplest approach that captures the essential physics underlying the
problem is low-dimensional dynamical modeling and analysis
\cite{Sugama_Horton_95,Ball_Dewar_Sugama_02,Ball_05,Kolesnikov_Krommes_05},
which can  provide a very economical tool to predict the
transition. However, the tradeoff with such  highly coarse-grained
modeling is that it necessarily whites out information, and may
therefore miss important physics and predict unphysical singular
behavior \cite{Ball_05}. Thus we require validation of the
low-dimensional modeling results by computational simulations of finer
models. 

The HW model \cite{Hasegawa:1983,Wakatani_Hasegawa_84} was developed
to investigate anomalous edge transport due to collisional drift waves,
and has been widely studied
\cite{Hasegawa_Wakatani_87,Horton_99,Pedersen_Michelson_Rasmussen_96,Camargo_Biskamp_Scott_95,Gang_Scott_Diamond_89}. It
includes the effects of inhomogeneous background density and parallel
electron dynamics described by Ohm's law.  The density gradient drives
the drift waves, which are destabilized by the parallel electron
resistivity. Convective nonlinearity regulates the linear growth of the
resistive drift wave instability, and a quasi-stationary state is
achieved where the resistive coupling balances the input. The HW model
is particularly simple yet includes the essential physics for studying
the self-consistent generation of turbulence and growth and decay of
coherent macroscopic structures such as zonal flows
\cite{Hasegawa_Wakatani_87}, even though it does not describe physics
that can be important in specific situations, such as magnetic
curvature, magnetic shear, and electromagnetic effects.

We emphasize that the parallel electron motion is important for
generation, stabilization, and destabilization of zonal flows. The
parallel electron response given by the generalized Ohm's law leads to
resistive coupling between the electrostatic potential and the density
fluctuations. In toroidal geometry this coupling does not act on the
flux-averaged parts \cite{Dorland_Hammett_93}, and in the original or
unmodified HW model we do not observe zonal flows. Modification of the
resistive coupling term, described in Sec. \ref{sec:model}, enables the
generation of zonal flows. This corresponds to the difference between
the ITG and the ETG (electron-temperature-gradient) cases discussed by
Jenko \emph{et al.} \cite{Jenko_Dorland_Kotsch_Rogers_00}, who found
that suppression of the secondary KH instability in the ETG case, due to
the adiabatic electron response, is removed in the ITG limit.

In Sec.~\ref{sec:model}, we describe the HW model and discuss the
treatment of parallel electron motions. Linear stability analysis of
the zero-flow background is also given to calculate transition points in
parameter space. Numerical simulation results are given in
Sec.~\ref{sec:simulation_result}. We carry out a systematic parameter
survey to locate the transition from a zonal-flow-dominated state to a
turbulent state. To examine the hypothesis that this transition may be
ascribed to the tertiary KH instability of the zonal flow, we study the
KH stability of the generated zonal flows in the HW model in
Sec.~\ref{sec:stability} and compare the KH stability threshold with the
transition boundary determined by simulation. Discussions and
conclusions are presented in Sec.~\ref{sec:conclusion}.

\section{Modified Hasegawa--Wakatani Model}
\label{sec:model}

The physical setting of the HW model may be considered as the edge
region of a tokamak plasma of nonuniform density $n_{0}=n_{0}(x)$ and in
a constant equilibrium magnetic field $\bm{B}=B_{0}\nabla z$. Following
the drift wave ordering \cite{Hasegawa_Mima_77}, the ion vorticity
$\zeta\equiv\nabla^{2}\varphi$ ($\varphi$ is the electrostatic
potential, $\nabla^{2}=\p^{2}/\p x^{2}+\p^{2}/\p y^{2}$ is the 2D
Laplacian) and the density fluctuations $n$ are governed by the
equations
\begin{align}
 \pdf{}{t} \zeta + \{\varphi,\zeta \}
  & = \alpha(\varphi-n) - D \nabla^{4}\zeta,
  \label{eq:hw_vorticity}\\
 \pdf{}{t} n + \{\varphi,n \}
 & = \alpha(\varphi-n) - \kappa \pdf{\varphi}{y} - D
 \nabla^{4} n,
 \label{eq:hw_density}
\end{align}
where $\{a,b\}\equiv (\p a/\p x) (\p b/\p y) -(\p a /\p y) (\p b/\p x)$
is the Poisson bracket, $D$ is the dissipation coefficient. The
background density is assumed to have an unchanging exponential profile:
$\kappa \equiv (\p/\p x) \ln n_{0}$. Electron parallel motion is
determined by Ohm's law with electron pressure
$p_{\mathrm{e}}=nT_{\mathrm{e}}$,
\begin{equation}
 j_{z}=-en v_{\mathrm{e},z}
  = - \frac{1}{\eta}\frac{\p}{\p z} 
  \left( \varphi- \frac{T_{\mathrm{e}}}{e} \ln n \right),
  \label{eq:current}
\end{equation}
assuming electron temperature $T_{\mathrm{e}}$ to be constant
(isothermal electron fluid).  This relation gives the coupling between
$\zeta$ and $n$ through the adiabaticity operator $\alpha \equiv
-T_{\mathrm{e}}/(\eta n_{0}\omega_{\mathrm{ci}} e^{2}) \p^{2}/\p
z^{2}$ appearing in Eqs.~(\ref{eq:hw_vorticity}) and
(\ref{eq:hw_density}).  In our 2D setting $\alpha$ becomes a
constant coefficient when acting on the drift wave components of
$\varphi$ and $n$ by the replacement $\p/\p z \rightarrow \imag
k_{z}$, where $2\pi/k_{z} = L_{\parallel} \gg L_{y}$ is a length
characteristic of the drift waves' phase variation along the field
lines.  However, for the zonal flow components, this resistive
coupling term must be treated carefully because zonal components of
fluctuations ($k_{y}=k_{z}=0$ modes) do not contribute to the parallel
current \cite{Dorland_Hammett_93}.  Recalling that turbulence in the tokamak
edge region, where there is strong magnetic shear, is considered here,
$k_{y}=0$ should always coincide with $k_{z}=0$ because any potential
fluctuation on the flux surface is neutralized by parallel electron
motion. Let us define zonal and non-zonal components of a variable $f$
as 
\begin{equation*}
 \textrm{zonal:}~
  \langle f \rangle = \frac{1}{L_{y}}\int f \diff y,~~~
  \textrm{non-zonal:}~
  \tilde{f} = f - \langle f \rangle,
  \label{eq:zonal_nonzonal}
\end{equation*}
where $L_{y}$ is the periodic length in $y$, and remove the contribution
by the zonal components in the resistive coupling term in
Eqs.~(\ref{eq:hw_vorticity}) and (\ref{eq:hw_density}). Subtraction of
the zonal components from the resistive coupling term $\alpha
(\varphi-n) \rightarrow \alpha(\tilde{\varphi}-\tilde{n})$ yields the
modified HW (MHW) equations, 
\begin{align}
 \pdf{}{t} \zeta + \{\varphi,\zeta \}
  & = \alpha(\tilde{\varphi}-\tilde{n}) - D \nabla^{4}\zeta, 
  \label{eq:mhw_vorticity}\\
 \pdf{}{t} n + \{\varphi,n \}
 & = \alpha(\tilde{\varphi}-\tilde{n}) - \kappa \pdf{\varphi}{y} - D
 \nabla^{4} n.
 \label{eq:mhw_density}
\end{align}
Evolutions of the zonal components can be extracted from
Eqs.~(\ref{eq:mhw_vorticity}) and (\ref{eq:mhw_density}) by averaging in
the $y$ direction:
\begin{equation*}
 \pdf{}{t} \langle f \rangle + \pdf{}{x}
  \left\langle f v_{x} \right\rangle
  = - D \frac{\p^{4}}{\p x^{4}} \langle f
  \rangle,~~ v_{x}\equiv -\pdf{\tilde{\varphi}}{y},
  \label{eq:mhw_zonal}
\end{equation*}
where $f$ stands for $\zeta$ and $n$.

Wakatani and Hasegawa found \cite{Wakatani_Hasegawa_84} that excitations
of waves having $k_{z}$ that maximizes the linear growth rate (for given
$k_{x}$ and $k_{y}$) are most likely to occur, since the plasma can
choose any parallel wavenumber ($k_{z}$).  Using the parallel wave
number of the maximum growth rate, $\alpha$ is given by $\alpha=4 k^2
k_y \kappa /(1+k^2)^2$.  This also gives $\alpha=0$ for the zonal mode. 

The MHW model spans two limits with respect to the adiabaticity
parameter $\alpha$. In the adiabatic limit $\alpha \rightarrow \infty$
(collisionless plasma), the non-zonal component of electron density obeys the
Boltzmann relation $\tilde{n}=n_{0}(x) \exp(\tilde{\varphi})$, and the
equations are reduced to the Hasegawa--Mima equation \cite{Hasegawa_Mima_77}. 
In the hydrodynamic limit $\alpha \rightarrow 0$, the equations are
decoupled. The vorticity is determined by the 2D Navier-Stokes equation,
and the density becomes a passive scalar. The advantage of our choice of
$\alpha$ as a free parameter is the capability for treating the  limits
in a unified manner.

The variables in Eqs.~(\ref{eq:mhw_vorticity}) and
(\ref{eq:mhw_density}) have been normalized by 
\begin{equation*}
 x/\rho_{\mathrm{s}} \rightarrow x, ~~
  \omega_{\mathrm{ci}} t \rightarrow t, ~~
  e \varphi / T_{\mathrm{e}} \rightarrow \varphi, ~~
  n_{1}/n_{0} \rightarrow n,
  \label{eq:normalization}
\end{equation*}
where $\rho_{\mathrm{s}}\equiv
 \sqrt{T_{\mathrm{e}}/m}\omega_{\mathrm{ci}}^{-1}$ is the ion sound
 Larmor radius ($v_{\mathrm{si}}\equiv\sqrt{T_{\mathrm{e}}/m}$ is the
 ion sound velocity in the cold ion limit), $n_{1}$ is the fluctuating
 part of the density.

In the adiabatic, ideal limit ($\alpha=\infty$, $D=0$) the MHW system
has two dynamical invariants, the energy $E$ and the potential enstrophy
$W$, 
\begin{equation}
 E = \frac{1}{2} \int (n^{2}+|\nabla \varphi|^{2}) \diff\bm{x},~~~
  W = \frac{1}{2} \int (n-\zeta)^{2} \diff \bm{x},
  \label{eq:energy_enstrophy}
\end{equation}
where $\diff\bm{x}=\diff x \diff y$, which constrain the fluid
motion. Conservation laws are given by
\begin{equation}
 \odf{E}{t} = \Gamma_{n} - D_{\alpha} - D_{E},~~~
  \odf{W}{t} = \Gamma_{n} - D_{W},
  \label{eq:energy_enstrophy_conservation}
\end{equation}
where fluxes and dissipations are given by
\begin{align*}
 \Gamma_{n} & = - \kappa \int \tilde{n}\pdf{\tilde{\varphi}}{y}
 \diff\bm{x},
 \\
 D_{\alpha} & = \alpha \int (\tilde{n}-\tilde{\varphi})^{2} \diff\bm{x},
 \\
 D_{E} & = D \int ( (\nabla^{2} n)^{2} + |\nabla \zeta|^{2} )
   \diff\bm{x},
 \\
 D_{W} & = D \int (\nabla^{2} n - \nabla^{2} \zeta)^{2}
  \diff\bm{x}.
\end{align*}
Unlike the Hasegawa--Mima model which is an energy-conserving system,
the MHW model has an energy source $\Gamma_{n}$. Due to the parallel
resistivity, $\tilde{n}$ and $\tilde{\varphi}$ can fluctuate out of phase
which produces non-zero $\Gamma_{n}$. The system can absorb free energy 
contained in the background density profile through the resistive drift
wave instability.

Note that the same conservation laws hold for the original, unmodified
original HW (OHW) model, Eqs.~(\ref{eq:hw_vorticity}) and
(\ref{eq:hw_density}), except that $D_{\alpha}$ is defined by both zonal 
and non-zonal components; $D_{\alpha}^{\textrm{OHW}}\equiv
\alpha\int(n-\varphi)^{2}\diff \bm{x}$. In the OHW model, the zonal
modes as well as the non-zonal modes suffer resistive dissipation.

\begin{figure}[htbp]
 \begin{center}
 \includegraphics[scale=.8]{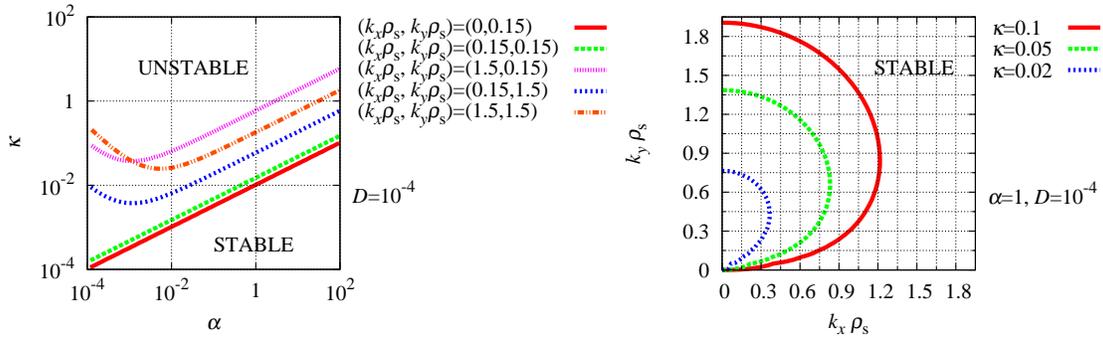}
 \caption{Primary stability boundary in $\alpha$-$\kappa$ plane and
  $k_{x}$-$k_{y}$ plane.}
 \label{fig:hw_stability}
 \end{center}
\end{figure}

We present the linear stability analysis for the zero background (the
primary instability). Beyond this stability threshold we expect 
excitation of drift waves. Since the zonal modes have linearly decaying
solutions, we only consider the form $\exp\imag (k_{x}x+k_{y}y-\omega
t)$ ($k_{y}\neq 0$). Linearization of Eqs.~(\ref{eq:mhw_vorticity})
and (\ref{eq:mhw_density}) around the zero equilibrium ($\varphi=n=0$)
yields the dispersion relation,
\begin{equation}
 \omega^{2} + \imag \omega (b+2 Dk^{4}) -
  \imag b \omega_{\ast} - \alpha D k^{2}(k^{2}+1) -
  D^{2} k^{8} = 0,
  \label{eq:hw_dispersion}
\end{equation}
where we defined $k^{2}=k_{x}^{2}+k_{y}^{2}$, $b\equiv\alpha
(1+k^{2})/k^{2}$, and the drift frequency $\omega_{\ast}\equiv k_{y}
\kappa/(1+k^{2})$. Solutions to the dispersion relation
(\ref{eq:hw_dispersion}) are given by 
\begin{align*}
 \omega_{\mathrm{r}} & = \pm \frac{b}{2}
 \left( 1 + \frac{16 \omega_{\ast}^{2}}{b^{2}} \right)^{\frac{1}{4}}
 \cos\frac{\theta}{2}, 
 \\
 \omega_{\mathrm{i}} & = - \frac{1}{2}
 \left[
 b+2Dk^{4} \mp
 b \left( 1 + \frac{16 \omega_{\ast}^{2}}{b^{2}} \right)^{\frac{1}{4}}
 \sin \frac{\theta}{2}
 \right],
\end{align*}
$\omega=\omega_{\mathrm{r}}+\imag \omega_{\mathrm{i}}$,
$\tan \theta = -4 \omega_{\ast}/b$. In the limit where $D=0$, it is
readily proved that one of the growth rates $\omega_{\mathrm{i}}$ is
positive if $b\omega_{\ast}$ is finite, thus unstable. However, there
exists a range of $D$ where the drift wave instability is
suppressed. The stability threshold is given by
 \begin{equation}
  b+2 D k^{4} \geq
   b \left( 1 + \frac{16\omega_{\ast}^{2}}{b^{2}} \right)^{\frac{1}{4}}
  \sin \frac{\theta}{2},
  \label{eq:hw_stability_condition}
 \end{equation}
 and is depicted in Fig.~\ref{fig:hw_stability}. The first unstable mode
 shown in the figure is the $(k_{x}\rho_{\mathrm{s}},
 k_{y}\rho_{\mathrm{s}})=(0,0.15)$ mode. Below this threshold, an
 initial perturbation damps out and nothing happens. If we choose the
 parameters in the region beyond the threshold, more than one mode
 starts to grow linearly until the nonlinear terms set in. The left
 panel shows how many modes are excited for given parameters. Most
 unstable modes are on $k_{x}=0$ axis.

\section{Simulation Results}
\label{sec:simulation_result}

The HW equations are solved in a doubly periodic square slab domain with
box size $L=2\pi/\Delta k$ where the lowest wavenumber $\Delta k = 0.15$
($L \sim 42$). The equations are discretized on $256\times 256$ grid 
points by the finite difference method. Arakawa's method is used for
evaluation of the Poisson bracket \cite{Arakawa_66}. The time stepping
algorithm is the third order explicit linear multistep method
\cite{Karniadakis_Israeli_Orszag_91}. We examine the effects of 
the parameters $\kappa$ and $\alpha$ on the nonlinearly saturated state,
and fix $D=10^{-4}$ throughout this paper.

\begin{figure}
 \begin{center}
  \includegraphics[scale=.9]{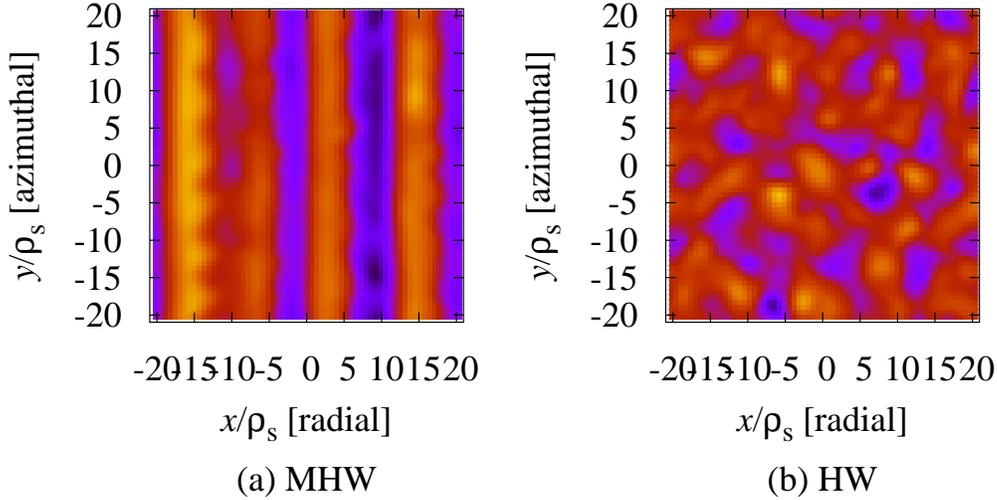}
  \caption{Contour plot of $\varphi$ in the saturated state. Zonally
  elongated structure of the electrostatic potential is clearly visible
  in the modified HW model (a), while isotropic vortices are generated
  in the HW model (b).}
  \label{fig:potential_contour}
 \end{center}
\end{figure}

\begin{figure}
 \begin{center}
  \includegraphics[scale=0.9]{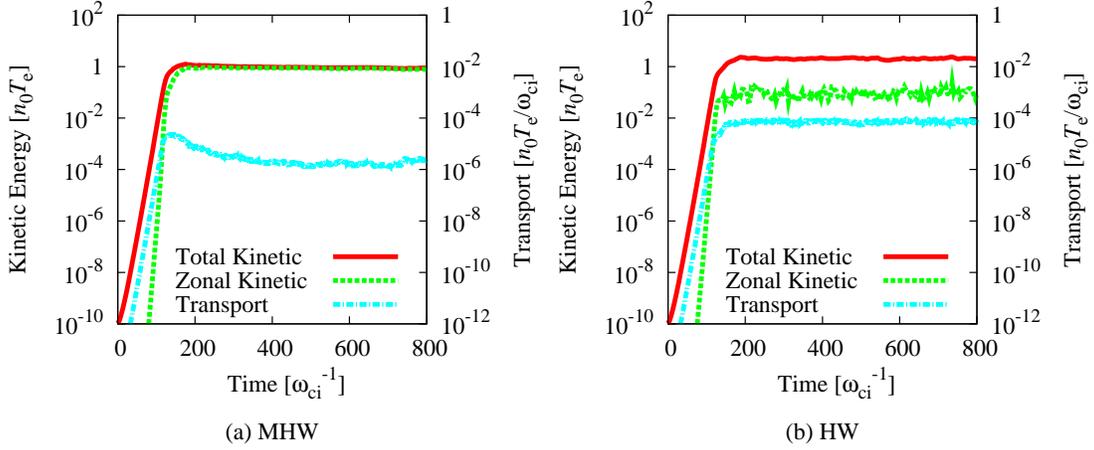}
  \caption{Time evolution plots of total kinetic energy, zonal flow kinetic
  energy and transport of MHW and HW models} 
  \label{fig:turbulence_suppression}
 \end{center}
\end{figure}

We start simulations by imposing small amplitude random
perturbations. The perturbations grow linearly in the initial phase and
generate drift waves, then the drift waves undergo secondary
instabilities which excite zonal flows until nonlinear saturation
occurs. In the saturated state, we observe that $\Gamma_{n}\simeq
D_{\alpha}\gg D_{E}, D_{W}$.  We compare the MHW and the OHW models by
showing the spatial behavior of the saturated electrostatic potential
in Fig.~\ref{fig:potential_contour}, and the time evolution of the
total kinetic energy, the zonal component of the kinetic energy, and
the cross-field transport $\Gamma_{n}$ in
Fig.~\ref{fig:turbulence_suppression}. From
Fig.~\ref{fig:potential_contour} we see that zonally elongated
structures of the electrostatic potential are generated in the MHW
model, while rather isotropic vortices are generated in the OHW model.
From Fig.~\ref{fig:turbulence_suppression} we see that growth of
the drift waves is not changed by the modification, but that in the
MHW model the zonal flows saturate at a higher amplitude (because the
modification removes the unphysical resistive dissipation of the zonal
modes). In fact, in the MHW model, the zonal flows carry nearly all
the kinetic energy in the final state --- they have absorbed nearly all
the energy from the drift waves. In both models, the cross-field
transport initially increases as the turbulent kinetic energy level
increases, but in the MHW model it begins to fall as zonal flows
absorb the drift wave energy. The build-up of the zonal flow in the
MHW model and the resulting transport suppression highlight the
importance of the difference between the MHW and the original HW model
in the nonlinear regime \cite{Numata_Ball_Dewar_07}.


\begin{figure}[htbp]
 \begin{center}
  \includegraphics[scale=0.8]{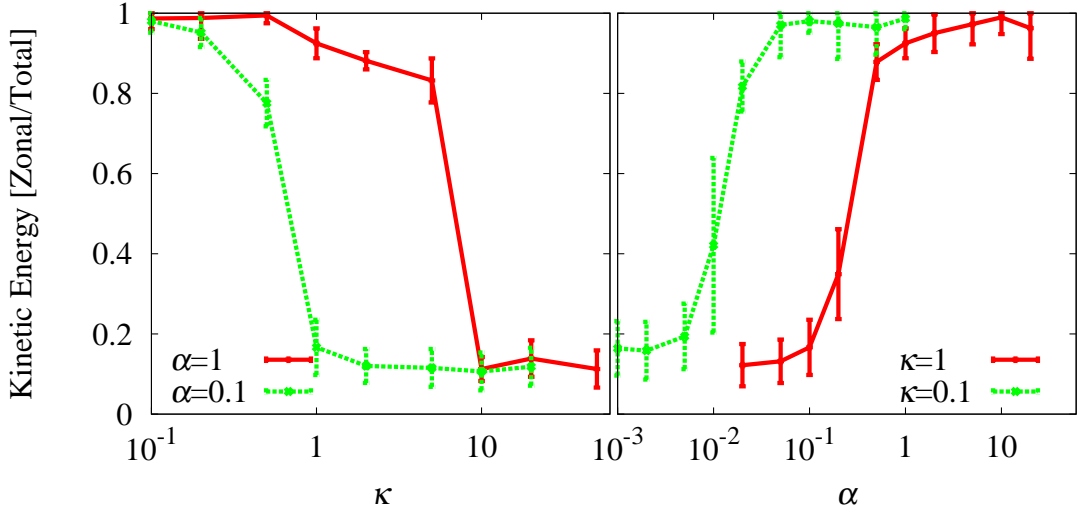}
 \caption{Parameter dependence of the zonal kinetic energy normalized by
  the total kinetic energy. Transitions from a zonal-flow-dominated
  state to a turbulence-dominated state occur.}
 \label{fig:bifurcation}
 \end{center}
\end{figure}

Let us show how the parameters $\kappa$ and $\alpha$ affect the
saturated state in the MHW model.  In Fig.~\ref{fig:bifurcation}, we
plot the ratio of the kinetic energy of the zonal flow
($F\equiv1/2\int(\p \langle\varphi\rangle/\p x)^{2}\diff \bm{x}$) to
the total kinetic energy ($E^{\mathrm{k}}
\equiv1/2\int|\nabla\varphi|^{2}\diff \bm{x}$) against $\kappa$ and
$\alpha$.  It is clearly seen that there are two types of saturated
states.  One is a zonal-flow-dominated state where turbulence is
almost completely suppressed, and the other is an isotropic
turbulence-dominated state.  The zonal-flow-dominated state suddenly
jumps to the turbulent state in a narrow range of the parameter space.
If we strongly drive the drift wave instability by increasing
$\kappa$, the system is likely to reach the turbulent state.  From the
dependence on $\alpha$, we can see that zonal flows are generated in
the adiabatic regime ($\alpha \gg 1$) while isotropic flows are
generated in the hydrodynamic regime ($\alpha\ll 1$).  These results are 
compatible with the properties of the Hasegawa--Mima model and of
hydrodynamic flows as discussed in the next section.

\begin{figure}
 \begin{center}
  \includegraphics[scale=0.8]{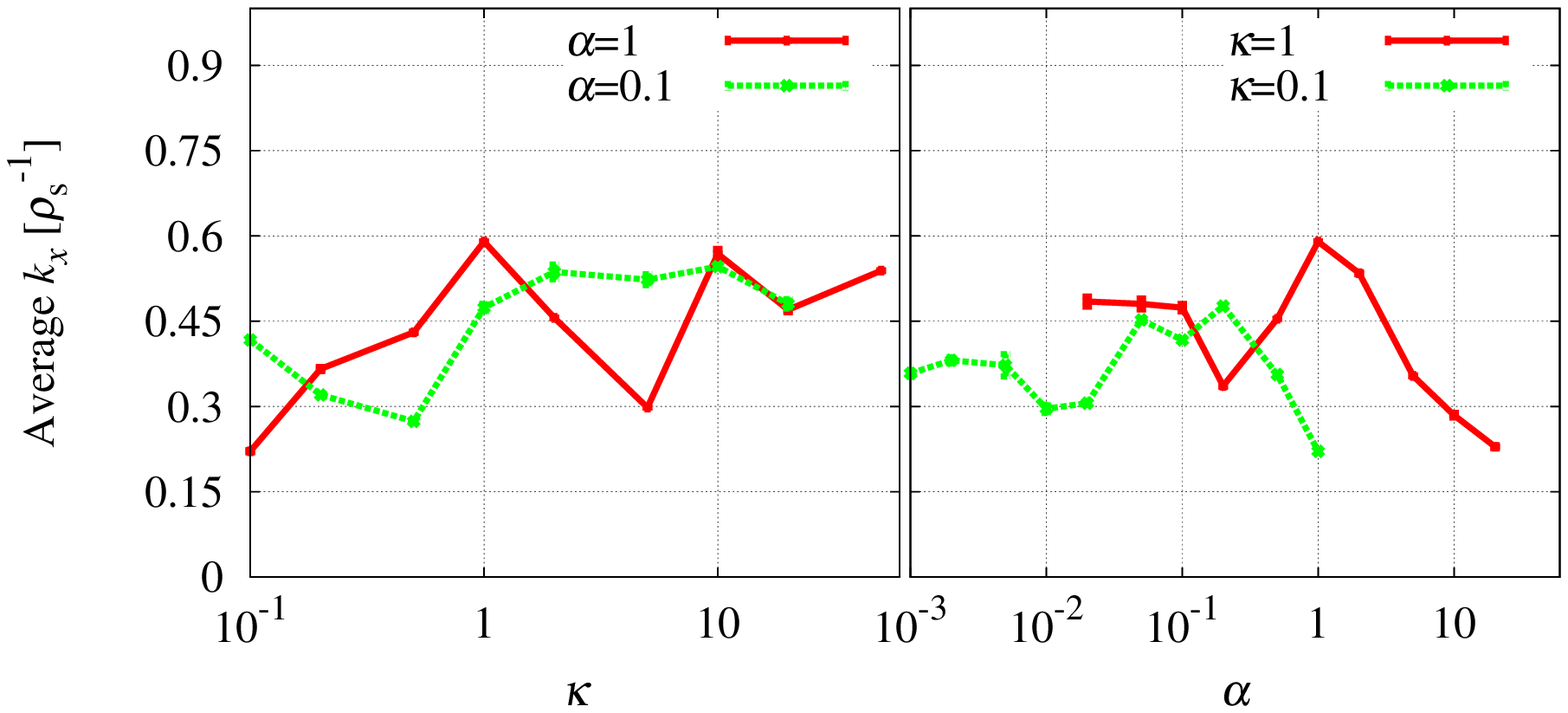}
  \caption{Average zonal flow wavenumber versus $\kappa$ and $\alpha$.}
  \label{fig:wavenumber}
 \end{center}
\end{figure}

\begin{figure}
 \begin{center}
  \includegraphics [scale=0.8]{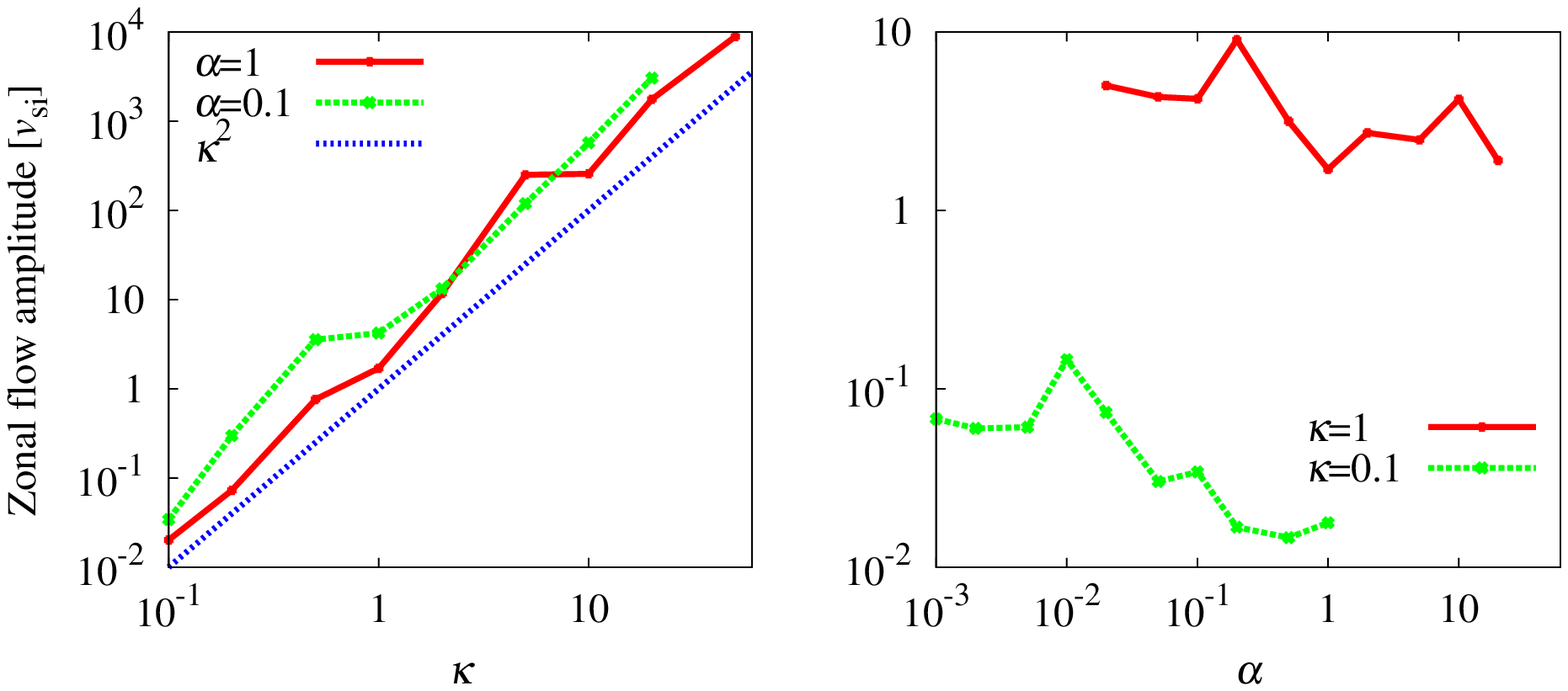}
  \caption{Zonal flow amplitude versus $\kappa$ and $\alpha$.}
  \label{fig:amplitude}
 \end{center}
\end{figure}

Let us assume that the generated zonal flows in the $y$ direction can be
expressed by a sinusoidal profile,
\begin{equation}
 V(x)= V_{0} \sin (\lambda x).
  \label{eq:flowprofile}
\end{equation}
The amplitude $V_{0}$ and wavenumber $\lambda=n_{\lambda}\pi/L$ are
determined from the simulation results. To estimate $\lambda$ we plot
the average wavenumber of the generated zonal flow
\begin{equation}
 \langle k_{x} \rangle = \frac{\int k_{x} {\cal E}^{k} (k_{x},k_{y}=0) \diff
  k_{x}} {\int {\cal E}^{k} (k_{x},k_{y}=0)\diff k_{x}}~~~~({\cal E}^{k}
  {\textrm{ is the kinetic energy spectrum}})
  \label{eq:average_wavenumber}
\end{equation}
in Fig.~\ref{fig:wavenumber}, and amplitude of the zonal flow in
Fig.~\ref{fig:amplitude}. The average wavenumbers are small and
rather insensitive to the parameters. This illustrates a feature of 2D
flows, which tend to generate large scale structures. The wavenumber of
a stable zonal flow is typically $0.3$ (corresponding to
$n_{\lambda}=4$). The amplitudes of zonal flows are roughly
proportional to $\kappa^{2}$ and are independent of $\alpha$.

\section{Stability of Zonal Flow}
\label{sec:stability}

We examine the stability of the zonal flows obtained from the numerical
simulations, and compare the stability threshold and the transition
point in this section. We consider the perturbation around the zonal
flow background. The electrostatic potential and the density are
decomposed as $\varphi=\varphi_{0}(x)+\hat{\varphi}(x)\exp\imag(k_y
y-\omega t)$, and $n=\hat{n}(x)\exp\imag(k_y y-\omega t)$ where $\diff 
\varphi_{0}/\diff x=V$ gives the background flow in the $y$
direction. By linearizing the MHW equations, we obtain an eigenvalue
equation containing the effect of $\kappa$ and $\alpha$,  
\begin{equation}
 \left[
 \odft{}{x} - k_{y}^{2} + \frac{k_{y}V''}{\omega-k_{y}V}
 - \frac{\imag \alpha}{\omega - k_{y} V + \imag\alpha}
 \left(1-\frac{k_{y}\kappa}{\omega-k_{y}V}
 \right)\right]\hat{\varphi}=0.
 \label{eq:eveq_hw}
\end{equation}
We neglect the viscosity. The density fluctuation is determined by
\begin{equation}
 \hat{n} = \frac{\imag \alpha + k_{y} \kappa}{\omega - k_{y}V + \imag
 \alpha} \hat{\varphi}.
 \label{eq:n_phi_relation}
\end{equation}
We solve the eigenvalue equation by the standard shooting method in the
domain ${\cal D}=\{x|-L/2\leq x\leq L/2\}$. The boundary is assumed to
be rigid $\hat{\varphi}(\pm L/2)=0$ for simplicity. 

\subsection{Hydrodynamic and adiabatic limit}
\label{sec:stability_limit}

Before going to the analysis of the HW case, we briefly
review the results in two limits: the hydrodynamic limit
($\alpha\rightarrow 0$) and the adiabatic limit
($\alpha\rightarrow\infty$).

In the $\alpha\rightarrow 0$ limit, we recover the Rayleigh eigenvalue
equation for neutral fluids,
\begin{equation}
 \left[
 \odft{}{x} - k_{y}^{2} + \frac{k_{y}V''}{\omega-k_{y}V} \right]
 \hat{\varphi}=0.
  \label{eq:eveq_hydrodynamic}
\end{equation}
The well-known Rayleigh's inflection point theorem demands existence
of an inflection point for the instability \cite{Rayleigh_1880}. The
necessary and sufficient condition is also known for this
case. Tollmien \cite{Tollmien_1935} showed the existence of a marginally
stable eigenfunction $\varphi_{\mathrm{s}}$ satisfying 
$\omega_{\mathrm{s}}/k_{\mathrm{s},0}=V(x_{\mathrm{s}})$ where
$x_{\mathrm{s}}$ is the inflection point. $\varphi_{\mathrm{s}}$ satisfies,
\begin{equation}
 \varphi_{\mathrm{s}}''+(\lambda^{2}-k_{\mathrm{s},0}^{2})\varphi_{\mathrm{s}}
  =0.
  \label{eq:marginal_hydrodynamic}
\end{equation}
The solution is given by
\begin{equation}
 \varphi_{\mathrm{s}} = \left\{
		\begin{matrix}
		 \sin(\frac{n\pi}{L}x) & (n: \textrm{even})\\
		 \cos(\frac{n\pi}{L}x) & (n: \textrm{odd})
		\end{matrix}
	       \right.,
 \label{eq:marginal_eigenfunction}
\end{equation}
and the critical wave number is
\begin{equation}
  k_{\mathrm{s},0} = \sqrt{\lambda^{2}-\left(\frac{n \pi}{L}\right)^{2}}
  ~~(n=\pm 1,\pm 2,\cdots).
  \label{eq:ks_hydrodynamic}
\end{equation}
If $\lambda>\pi/L$, the marginally stable wave number $k_{\mathrm{s},0}$
exists. It should be noted that Tollmien does not exclude the possibility
that the marginally stable mode is isolated. However, perturbation
analysis around the marginally mode shows the existence of solutions
smoothly connected to the marginal solution \cite{Lin_1945_II,Drazin_81}.

A similar analysis can be applied to the adiabatic limit,
\begin{equation}
 \left[
 \odft{}{x} - (k_{y}^{2}+1) + \frac{k_{y}(V''+\kappa)}{\omega-k_{y}V}
 \right]\hat{\varphi}=0
  \label{eq:eveq_adiabatic}
\end{equation}
if $\kappa=0$. The marginally stable eigenfunction satisfies,
\begin{equation}
 \varphi_{\mathrm{s}}''+(\lambda^{2}-k_{\mathrm{s},\infty}^{2}-1)
  \varphi_{\mathrm{s}} =0. 
\end{equation}
The solution is identical with the previous case, but the critical wave
number is slightly modified to
\begin{equation}
 k_{\mathrm{s},\infty} = \sqrt{\lambda^{2}-\left(\frac{n\pi}{L}\right)^{2}-1}.
  \label{eq:ks_adiabatic}
\end{equation} 
The necessary and sufficient condition of the flow shear for instability
is $\lambda^{2}>(\pi/L)^{2}+1$.

We can judge the stability by finding the critical wavenumber. We
consider the flow given by Eq.~(\ref{eq:flowprofile}) with
$\lambda=0.3$. The critical wavenumber exists only in the hydrodynamic
limit for the given profile. On the other hand, the given flow is stable
in the adiabatic limit. The difference of the two stability conditions
(\ref{eq:ks_hydrodynamic}) and (\ref{eq:ks_adiabatic}) comes not from
$\kappa$ but from the strong coupling between $\varphi$ and $n$, and
reflects the stabilizing effect of adiabatic parallel electron motion.

\begin{center}
 \begin{figure}
  \includegraphics[scale=0.6]{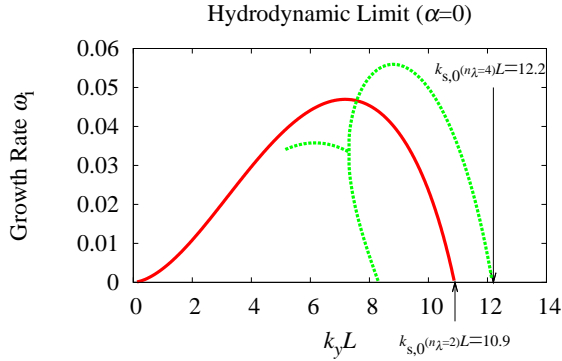}
  \caption{Growth rates for $\lambda=0.3$ flow in hydrodynamic limit 
  as described in the text.}
  \label{fig:ev_hydrodynamic}
 \end{figure}
\end{center}

Figure~\ref{fig:ev_hydrodynamic} shows the imaginary parts of the
eigenvalues for $n_{\lambda}=4$ case in the hydrodynamic limit. The
eigenvalues are pure imaginary in this limit because of antisymmetry of
the flow [$V(x)=-V(-x)$]. Another property in this limit is the scale 
invariance. The eigenvalues do not depend on $L$ and $\lambda$, but
are determined by $n_{\lambda}$.

We set $V_{0}=1$. Or, in  other words, $V_{0}$ is normalized out by
considering $\omega/V_{0} \rightarrow \omega$. The eigenvalue problem of
the given flow profile with $n_{\lambda}=4$ has the same eigenvalues as
that of the flow with $n_{\lambda}=2$ in the half domain (solid line).
The critical wavenumber for this curve is given by
$k_{\mathrm{s},0}(n_{\lambda}=2) \sim 0.26$. In addition, we find
another branch of solutions (broken line) which continue to exist until
$k_{y}<k_{\mathrm{s},0}(n_{\lambda}=4) \sim 0.29$.

Next, let us consider the effect of $\kappa$. Since the critical wavenumber
does not exist for the profile with $\lambda=0.3$, we examine a
profile having stronger flow shear by setting $L=5$, and take
$n_{\lambda}=2$ for simplicity. In this setting the marginal wavenumber
exists ($k_{\mathrm{s},\infty}\sim2.14$).
\begin{center}
 \begin{figure}
  \includegraphics[scale=0.8]{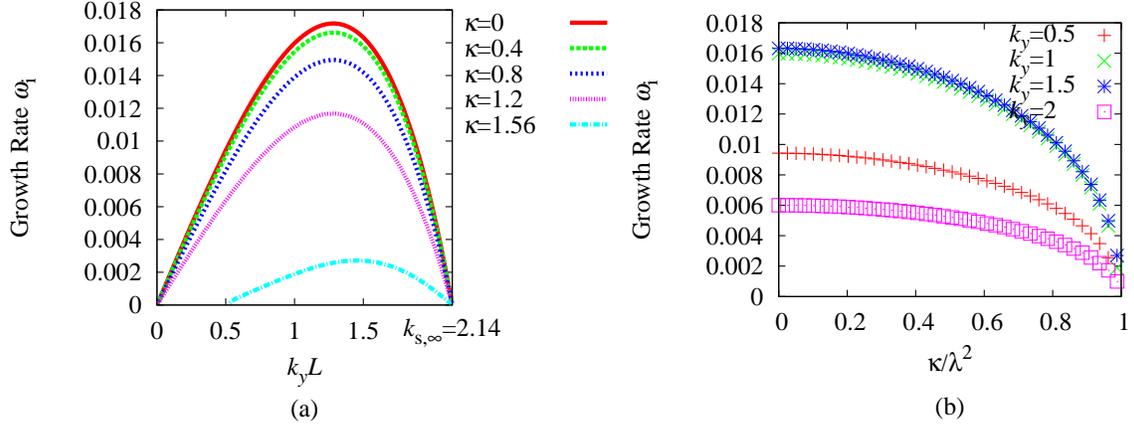}
  \caption{Growth rate in adiabatic limit ($L=5$, $n_{\lambda}=2$). (a)
  $k_{y}$ dependence, (b) $\kappa$ dependence.}
  \label{fig:ev_adiabatic}
 \end{figure}
\end{center}
Figure.~\ref{fig:ev_adiabatic} shows the eigenvalues obtained in the
adiabatic limit for $L=5$ and $\lambda=2\pi/L$. $\kappa$ is also
normalized by $\kappa/V_{0} \rightarrow \kappa$. $k_{\mathrm{s},\infty}$
seems independent of $\kappa$. Thus the same stability condition still
holds for finite, but not too large,  $\kappa$. As
we see from the figure, the growth rate $\omega_{\mathrm{i}}$ decreases
with increasing $\kappa$ and disappears for large $\kappa$ even though 
$k_{\mathrm{s},\infty}$ exists. We need another condition for
$\kappa$. Multiplying (\ref{eq:eveq_adiabatic}) by complex conjugate of
$\varphi$ and integrating over the domain, we obtain
\begin{equation}
 \omega_{\mathrm{i}} \int_{\cal D}
  \frac{k_{y}(V''+\kappa)}{|\omega-k_{y}V|^{2}} \diff x=0.
  \label{eq:condition_kappa_adiabatic}
\end{equation}
If $\omega_{\mathrm{i}}\neq 0$, $V''+\kappa=0$ must be satisfied
somewhere in the domain \cite{Kuo_1949}. Applying this condition to 
our assumed flow profile, we obtain the condition $\kappa<\lambda^{2}$
for the instability. This gives only a necessary condition for the
instability, but provides a good estimate [Fig.~\ref{fig:ev_adiabatic}
(b)].

If we find the eigenvalue $\omega$ and the corresponding eigenfunction
$\varphi$, the complex conjugate of $\omega$ is also an eigenvalue and the
corresponding eigenfunction is given by the complex conjugate of $\varphi$.
Thus, we can always restrict our quest for eigenvalues in the upper
half plane of the complex $\omega$ plane without loss of
generality. This greatly simplifies the situation because we can neglect
the continuous spectrum on the real $\omega$ axis. 

\subsection{Hasegawa--Wakatani case (intermediate value of $\alpha$)}
\label{sec:stability_hw}

Unlike the previous cases, the complex conjugate of an eigenvalue is
not a eigenvalue if we include finite $\alpha$. In this case we must
solve for negative $\omega_{\mathrm{i}}$ as well. Moreover, there exist
two continuous spectra in this case: 
\begin{equation}
 \omega = k_{y}V,\, k_{y}V - \imag \alpha
  ~~~ {\mathrm {where}}~|V|\leq V_{0}.
\end{equation}
Both represent convective transport due to the background flow.  One
of them is damped by the resistivity.  These continua may interact
with the point spectrum.  Thus the situation is much more complicated
in the intermediate $\alpha$ case compared with the adiabatic and
hydrodynamic limits.

\begin{center}
 \begin{figure}
  \includegraphics[scale=0.8]{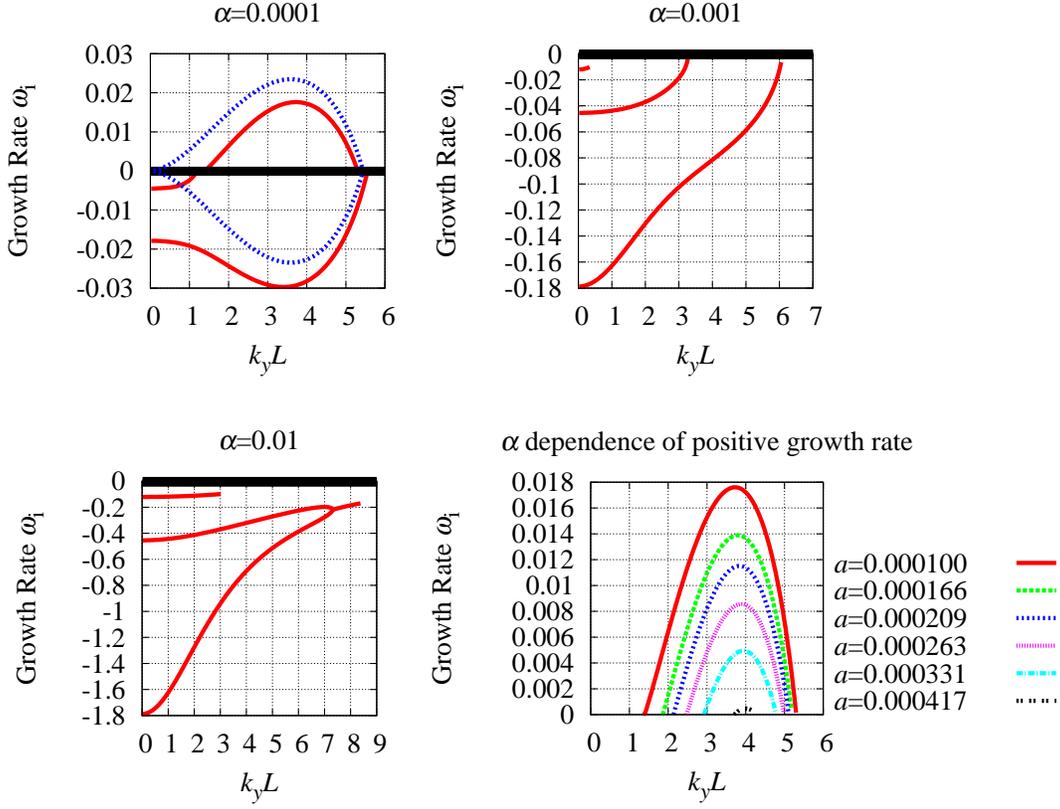}
  \caption{Growth rates for HW case as described in the text.}
  \label{fig:ev_hw}
 \end{figure}
\end{center}

First, we show the effect of $\alpha$ and neglect effect of $\kappa$. We
consider $n_{\lambda}=2$ for simplicity. Figure~\ref{fig:ev_hw} shows
the imaginary parts of the eigenvalues. Three different $\alpha$ cases,
and the $\alpha$ dependence of the positive branches, are shown.
The continuous spectra are shown by thick solid lines.  In the
$\alpha=0.0001$ case, two branches from the $\alpha\rightarrow 0$ case
(dotted line) are also shown for reference, so that it is seen that
$\omega_{\mathrm{i}}$ is slightly shifted downwards for finite
$\alpha$.  As $k_{y}L$ decreases, the upper (unstable) branch
intersects the continuous spectrum at marginal stability, and there
exists a gap (interval in $k_{y}L$) occupied by the two continuous
spectra before this branch continues as a stable mode.  The
eigenfunctions belonging to the eigenvalues in the point spectrum
close to this gap become singular.

For increasing $\alpha$, we observe the positive eigenvalues disappear
at $\alpha \approx 0.000417$.  In addition to the two stable branches
seen at $\alpha=0.0001$, at $\alpha=0.001$ another stable branch has
appeared in the small $k_{y}$ region.  By further increase of $\alpha$
we find that the lower two branches merge. Beyond this merging point,
finite real parts appear, and the eigenmode starts to travel in the $y$
direction.

Our concern is to determine the stability threshold in the
$\alpha$-$\kappa$ plane. Next, we consider the effect of $\kappa$ in
addition to $\alpha$. Since $\kappa$ always appears in the form of
$\kappa \alpha$ and $\alpha$ is small in the vicinity of the threshold,
the effect of $\kappa$ is rather minor. $\kappa$ does not significantly
affect the behavior of the eigenvalues except that $\kappa$ controls the
amplitude of flow. As we stated earlier, the parameters are normalized
by $V_{0}$, $\kappa/V_{0} \rightarrow \kappa$, $\alpha/V_{0} \rightarrow
\alpha$, in the shooting calculation, where $V_{0}$ is proportional to
$\kappa^{2}$.

\begin{center}
 \begin{figure}
  \includegraphics[scale=0.6]{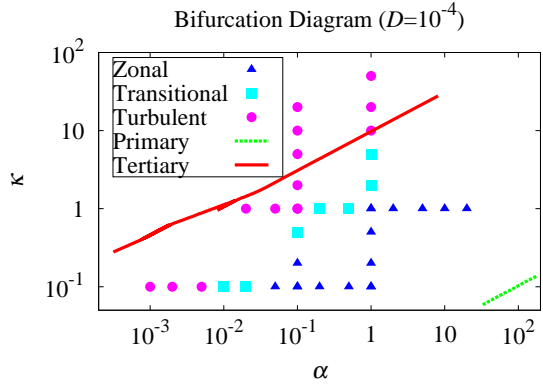}
  \caption{Bifurcation diagram showing the correlation between the
  linearized stability estimates described in the text and the regimes
  observed in our turbulence simulations.}
  \label{fig:bifucation_2d}
 \end{figure}
\end{center}

Finally, we summarize the shooting calculation by showing the
bifurcation diagram in $\alpha$-$\kappa$ plane together with the
numerically obtained results. The only excitable mode that can be
resolved in the numerical simulation is the $k_{y}=0.15$ mode, which
is the first unstable mode of the primary instability (see
Sec.~\ref{sec:model}).  In Fig.~\ref{fig:bifucation_2d}, we show the
stability threshold of $k_{y}=0.15$ mode for the primary instability
(resistive drift wave instability) and the tertiary instability
(KH instability). Each mark in the figure denotes a numerically
obtained saturated state: $\blacktriangle$, $\blacksquare$, $\bullet$
represent respectively the zonal-flow-dominated, transitional, and
turbulence-dominated states.  In these states zonal flows contain more
than 90\%, 20-90\%, and less than 20\% of the total kinetic energy,
respectively.  The qualitative tendency of the thresholds in the
bifurcation diagram shows agreement between the numerical simulations
and the KH analysis, i.e. increasing $\alpha$ ($\kappa$) is stabilizing
(destabilizing).  Zonal-flow-dominated states are observed in between
the primary and the tertiary instability thresholds.  The emergence of a
turbulent state is shifted from the primary threshold to the tertiary
threshold due to the turbulence suppression effect of the zonal flow,
which is analogous to the Dimits shift observed in ITG turbulence. 

The reasons for the quantitative discrepancy between the boundary of the
zonal and the turbulent states may be because of the simplification made  
in the KH analysis; the simplified flow profile, the boundary condition
and viscosity may also affect the results.

\section{Conclusion}
\label{sec:conclusion}

In summary, we have analyzed bifurcation phenomena in two-dimensional
resistive drift wave turbulence.  First, we have performed numerical
simulations of the modified HW model to study bifurcation structures in
a two-parameter ($\alpha$-$\kappa$) space. We have shown that, in the
MHW model, zonal flows are self-organized and suppress turbulence and
turbulent transport  over a range of parameters beyond the linear
stability threshold for resistive drift waves.  By performing a
systematic parameter survey, we have found that such
zonal-flow-dominated states suddenly disappear as a threshold is
crossed, being replaced by a turbulence-dominated state.

The threshold of the onset of turbulence has been compared with the
linear stability threshold of an assumed laminar zonal flow profile.
Simple theoretical predictions in limiting cases explain the
qualitative tendency of the stability of the zonal flow.  $\kappa$
determines the amplitude of the zonal flows, thus, large $\kappa$
destabilizes the zonal flows.  On the other hand, the adiabatic
response of parallel electrons given by $\alpha$ stabilizes them.
Numerical analysis of the eigenvalue problem determining the stability
of the assumed zonal flow profile in the HW model also confirms this
trend.  The constructed bifurcation diagram in the $\alpha$-$\kappa$
plane for the HW model confirms the scenario of the onset of turbulence
in the drift wave/zonal flow system being due to the disruption of zonal
flows by KH instability. 

The HW model considered here is a particularly simple
model, but includes the essential physics of interactions between
turbulence and coherent structures.  This system exhibits many other
interesting phenomena, but in this paper we have focused on the
effect of the linear driving term $\kappa$ and the parallel electron
response $\alpha$ (including the resistivity).  To do so,
we set the viscosity very small.  In this case the zonal flow survives
for a very long time.  However, when the viscosity comes into play,
the zonal flows are damped rapidly, and the turbulence grows again 
until zonal flows can be nonlinearly excited and the cycle repeats.
Thus the system exhibits predator-prey oscillatory behavior.

\section{Acknowledgments}

The authors would like to acknowledge B.~D. Scott for providing the
simulation code used in this work. We thank P.~N. Guzdar, W. Dorland,
C. Tebaldi and J.~A. Krommes for useful discussions. This work is
supported by the Australian Research Council.


\begin{thebibliography}{31}
\expandafter\ifx\csname natexlab\endcsname\relax\def\natexlab#1{#1}\fi
\expandafter\ifx\csname bibnamefont\endcsname\relax
  \def\bibnamefont#1{#1}\fi
\expandafter\ifx\csname bibfnamefont\endcsname\relax
  \def\bibfnamefont#1{#1}\fi
\expandafter\ifx\csname citenamefont\endcsname\relax
  \def\citenamefont#1{#1}\fi
\expandafter\ifx\csname url\endcsname\relax
  \def\url#1{\texttt{#1}}\fi
\expandafter\ifx\csname urlprefix\endcsname\relax\def\urlprefix{URL }\fi
\providecommand{\bibinfo}[2]{#2}
\providecommand{\eprint}[2][]{\url{#2}}

\bibitem[{\citenamefont{Diamond et~al.}(2005)\citenamefont{Diamond, Itoh, Itoh,
  and Hahm}}]{Diamond-Itoh-Itoh-Hahm_05}
\bibinfo{author}{\bibfnamefont{P.~H.} \bibnamefont{Diamond}},
  \bibinfo{author}{\bibfnamefont{S.-I.} \bibnamefont{Itoh}},
  \bibinfo{author}{\bibfnamefont{K.}~\bibnamefont{Itoh}}, \bibnamefont{and}
  \bibinfo{author}{\bibfnamefont{T.~S.} \bibnamefont{Hahm}},
  \bibinfo{journal}{Plasma Phys. Control. Fusion}
  \textbf{\bibinfo{volume}{47}}, \bibinfo{pages}{R35} (\bibinfo{year}{2005}).

\bibitem[{tra()}]{transition_by_external_flow}
\bibinfo{note}{The \emph{L--H} transition may also be induced by externally
  generated flows, for example by edge biasing [see R.~J. Taylor, M.~L. Brown,
  B.~D. Fried {\it et al.}, Phys. Rev. Lett. {\bf 63}, 2365 (1989)], but our
  interest here is focused on internally or self-generated shear flows.}

\bibitem[{\citenamefont{Hasegawa and Wakatani}(1983)}]{Hasegawa:1983}
\bibinfo{author}{\bibfnamefont{A.}~\bibnamefont{Hasegawa}} \bibnamefont{and}
  \bibinfo{author}{\bibfnamefont{M.}~\bibnamefont{Wakatani}},
  \bibinfo{journal}{Phys. Rev. Lett.} \textbf{\bibinfo{volume}{50}},
  \bibinfo{pages}{682} (\bibinfo{year}{1983}).

\bibitem[{\citenamefont{Wakatani and Hasegawa}(1984)}]{Wakatani_Hasegawa_84}
\bibinfo{author}{\bibfnamefont{M.}~\bibnamefont{Wakatani}} \bibnamefont{and}
  \bibinfo{author}{\bibfnamefont{A.}~\bibnamefont{Hasegawa}},
  \bibinfo{journal}{Phys. Fluids} \textbf{\bibinfo{volume}{27}},
  \bibinfo{pages}{611} (\bibinfo{year}{1984}).

\bibitem[{\citenamefont{Dimits et~al.}(2000)\citenamefont{Dimits, Bateman,
  Beer, Cohen, Dorland, Hammett, Kim, Kinsey, Kotschenreuther, Kritz
  et~al.}}]{Dimits_etal_00}
\bibinfo{author}{\bibfnamefont{A.~M.} \bibnamefont{Dimits}},
  \bibinfo{author}{\bibfnamefont{G.}~\bibnamefont{Bateman}},
  \bibinfo{author}{\bibfnamefont{M.~A.} \bibnamefont{Beer}},
  \bibinfo{author}{\bibfnamefont{B.~I.} \bibnamefont{Cohen}},
  \bibinfo{author}{\bibfnamefont{W.}~\bibnamefont{Dorland}},
  \bibinfo{author}{\bibfnamefont{G.~W.} \bibnamefont{Hammett}},
  \bibinfo{author}{\bibfnamefont{C.}~\bibnamefont{Kim}},
  \bibinfo{author}{\bibfnamefont{J.~E.} \bibnamefont{Kinsey}},
  \bibinfo{author}{\bibfnamefont{M.}~\bibnamefont{Kotschenreuther}},
  \bibinfo{author}{\bibfnamefont{A.~H.} \bibnamefont{Kritz}},
  \bibnamefont{et~al.}, \bibinfo{journal}{Phys. Plasmas}
  \textbf{\bibinfo{volume}{7}}, \bibinfo{pages}{969} (\bibinfo{year}{2000}).

\bibitem[{\citenamefont{Biskamp and Kaifen}(1985)}]{Biskamp:1985}
\bibinfo{author}{\bibfnamefont{D.}~\bibnamefont{Biskamp}} \bibnamefont{and}
  \bibinfo{author}{\bibfnamefont{H.}~\bibnamefont{Kaifen}},
  \bibinfo{journal}{Phys. Fluids} \textbf{\bibinfo{volume}{28}},
  \bibinfo{pages}{2172} (\bibinfo{year}{1985}).

\bibitem[{\citenamefont{Klinger et~al.}(1997)\citenamefont{Klinger, Latten,
  Piel, Bonhomme, Pierre, and Dudok~de Wit}}]{Klinger:1997}
\bibinfo{author}{\bibfnamefont{T.}~\bibnamefont{Klinger}},
  \bibinfo{author}{\bibfnamefont{A.}~\bibnamefont{Latten}},
  \bibinfo{author}{\bibfnamefont{A.}~\bibnamefont{Piel}},
  \bibinfo{author}{\bibfnamefont{G.}~\bibnamefont{Bonhomme}},
  \bibinfo{author}{\bibfnamefont{T.}~\bibnamefont{Pierre}}, \bibnamefont{and}
  \bibinfo{author}{\bibfnamefont{T.}~\bibnamefont{Dudok~de Wit}},
  \bibinfo{journal}{Phys. Rev. Lett.} \textbf{\bibinfo{volume}{79}},
  \bibinfo{pages}{3913} (\bibinfo{year}{1997}).

\bibitem[{\citenamefont{Ruelle and Takens}(1971)}]{Ruelle:1971}
\bibinfo{author}{\bibfnamefont{D.}~\bibnamefont{Ruelle}} \bibnamefont{and}
  \bibinfo{author}{\bibfnamefont{F.}~\bibnamefont{Takens}},
  \bibinfo{journal}{Commun. Math. Phys.} \textbf{\bibinfo{volume}{20}},
  \bibinfo{pages}{167} (\bibinfo{year}{1971}).

\bibitem[{\citenamefont{Guzdar et~al.}(2001)\citenamefont{Guzdar, Kleva, and
  Chen}}]{Guzdar_Kleva_Chen_01}
\bibinfo{author}{\bibfnamefont{P.~N.} \bibnamefont{Guzdar}},
  \bibinfo{author}{\bibfnamefont{R.~G.} \bibnamefont{Kleva}}, \bibnamefont{and}
  \bibinfo{author}{\bibfnamefont{L.}~\bibnamefont{Chen}},
  \bibinfo{journal}{Phys. Plasmas} \textbf{\bibinfo{volume}{8}},
  \bibinfo{pages}{459} (\bibinfo{year}{2001}).

\bibitem[{\citenamefont{Dewar and Abdullatif}(2007)}]{Dewar_Abdullatif_07}
\bibinfo{author}{\bibfnamefont{R.~L.} \bibnamefont{Dewar}} \bibnamefont{and}
  \bibinfo{author}{\bibfnamefont{R.~F.} \bibnamefont{Abdullatif}}, in
  \emph{\bibinfo{booktitle}{Proceedings of the CSIRO/COSNet Workshop on
  Turbulence and Coherent Structures, Canberra, Australia, 10-13 January
  2006}}, edited by \bibinfo{editor}{\bibfnamefont{J.~P.} \bibnamefont{Denier}}
  \bibnamefont{and} \bibinfo{editor}{\bibfnamefont{J.~S.}
  \bibnamefont{Frederiksen}} (\bibinfo{publisher}{World Scientific},
  \bibinfo{address}{Singapore}, \bibinfo{year}{2007}), vol.~\bibinfo{volume}{6}
  of \emph{\bibinfo{series}{World Scientific Lecture Notes in Complex
  Systems}}, pp. \bibinfo{pages}{415--430}.

\bibitem[{\citenamefont{Rogers et~al.}(2000)\citenamefont{Rogers, Dorland, and
  Kotschenreuther}}]{Rogers_Dorland_Kotsch_00}
\bibinfo{author}{\bibfnamefont{B.~N.} \bibnamefont{Rogers}},
  \bibinfo{author}{\bibfnamefont{W.}~\bibnamefont{Dorland}}, \bibnamefont{and}
  \bibinfo{author}{\bibfnamefont{M.}~\bibnamefont{Kotschenreuther}},
  \bibinfo{journal}{Phys. Rev. Lett.} \textbf{\bibinfo{volume}{85}},
  \bibinfo{pages}{5336} (\bibinfo{year}{2000}).

\bibitem[{\citenamefont{Jenko et~al.}(2000)\citenamefont{Jenko, Dorland,
  Kotschenreuther, and Rogers}}]{Jenko_Dorland_Kotsch_Rogers_00}
\bibinfo{author}{\bibfnamefont{F.}~\bibnamefont{Jenko}},
  \bibinfo{author}{\bibfnamefont{W.}~\bibnamefont{Dorland}},
  \bibinfo{author}{\bibfnamefont{M.}~\bibnamefont{Kotschenreuther}},
  \bibnamefont{and} \bibinfo{author}{\bibfnamefont{B.~N.}
  \bibnamefont{Rogers}}, \bibinfo{journal}{Phys. Plasmas}
  \textbf{\bibinfo{volume}{7}}, \bibinfo{pages}{1904} (\bibinfo{year}{2000}).

\bibitem[{\citenamefont{Sugama and Horton}(1995)}]{Sugama_Horton_95}
\bibinfo{author}{\bibfnamefont{H.}~\bibnamefont{Sugama}} \bibnamefont{and}
  \bibinfo{author}{\bibfnamefont{W.}~\bibnamefont{Horton}},
  \bibinfo{journal}{Plasma Phys. Control. Fusion}
  \textbf{\bibinfo{volume}{37}}, \bibinfo{pages}{345} (\bibinfo{year}{1995}).

\bibitem[{\citenamefont{Ball et~al.}(2002)\citenamefont{Ball, L.Dewar, and
  Sugama}}]{Ball_Dewar_Sugama_02}
\bibinfo{author}{\bibfnamefont{R.}~\bibnamefont{Ball}},
  \bibinfo{author}{\bibfnamefont{R.}~\bibnamefont{L.Dewar}}, \bibnamefont{and}
  \bibinfo{author}{\bibfnamefont{H.}~\bibnamefont{Sugama}},
  \bibinfo{journal}{Phys. Rev. E} \textbf{\bibinfo{volume}{66}},
  \bibinfo{pages}{066408} (\bibinfo{year}{2002}).

\bibitem[{\citenamefont{Ball}(2005)}]{Ball_05}
\bibinfo{author}{\bibfnamefont{R.}~\bibnamefont{Ball}}, \bibinfo{journal}{Phys.
  Plasmas} \textbf{\bibinfo{volume}{12}}, \bibinfo{pages}{090904}
  (\bibinfo{year}{2005}).

\bibitem[{\citenamefont{Kolesnikov and Krommes}(2005)}]{Kolesnikov_Krommes_05}
\bibinfo{author}{\bibfnamefont{R.~A.} \bibnamefont{Kolesnikov}}
  \bibnamefont{and} \bibinfo{author}{\bibfnamefont{J.~A.}
  \bibnamefont{Krommes}}, \bibinfo{journal}{Phys. Plasmas}
  \textbf{\bibinfo{volume}{12}}, \bibinfo{pages}{122302}
  (\bibinfo{year}{2005}).

\bibitem[{\citenamefont{Hasegawa and Wakatani}(1987)}]{Hasegawa_Wakatani_87}
\bibinfo{author}{\bibfnamefont{A.}~\bibnamefont{Hasegawa}} \bibnamefont{and}
  \bibinfo{author}{\bibfnamefont{M.}~\bibnamefont{Wakatani}},
  \bibinfo{journal}{Phys. Rev. Lett.} \textbf{\bibinfo{volume}{59}},
  \bibinfo{pages}{1581} (\bibinfo{year}{1987}).

\bibitem[{\citenamefont{Horton}(1999)}]{Horton_99}
\bibinfo{author}{\bibfnamefont{W.}~\bibnamefont{Horton}},
  \bibinfo{journal}{Rev. Mod. Phys.} \textbf{\bibinfo{volume}{71}},
  \bibinfo{pages}{735} (\bibinfo{year}{1999}).

\bibitem[{\citenamefont{Pedersen et~al.}(1996)\citenamefont{Pedersen,
  Michelsen, and Rasmussen}}]{Pedersen_Michelson_Rasmussen_96}
\bibinfo{author}{\bibfnamefont{T.~S.} \bibnamefont{Pedersen}},
  \bibinfo{author}{\bibfnamefont{P.~K.} \bibnamefont{Michelsen}},
  \bibnamefont{and} \bibinfo{author}{\bibfnamefont{J.~J.}
  \bibnamefont{Rasmussen}}, \bibinfo{journal}{Plasma Phys. Control. Fusion}
  \textbf{\bibinfo{volume}{38}}, \bibinfo{pages}{2143} (\bibinfo{year}{1996}).

\bibitem[{\citenamefont{Camargo et~al.}(1995)\citenamefont{Camargo, Biskamp,
  and Scott}}]{Camargo_Biskamp_Scott_95}
\bibinfo{author}{\bibfnamefont{S.~J.} \bibnamefont{Camargo}},
  \bibinfo{author}{\bibfnamefont{D.}~\bibnamefont{Biskamp}}, \bibnamefont{and}
  \bibinfo{author}{\bibfnamefont{B.~D.} \bibnamefont{Scott}},
  \bibinfo{journal}{Phys. Plasmas} \textbf{\bibinfo{volume}{2}},
  \bibinfo{pages}{48} (\bibinfo{year}{1995}).

\bibitem[{\citenamefont{Gang et~al.}(1989)\citenamefont{Gang, Scott, and
  Diamond}}]{Gang_Scott_Diamond_89}
\bibinfo{author}{\bibfnamefont{F.~Y.} \bibnamefont{Gang}},
  \bibinfo{author}{\bibfnamefont{B.~D.} \bibnamefont{Scott}}, \bibnamefont{and}
  \bibinfo{author}{\bibfnamefont{P.~H.} \bibnamefont{Diamond}},
  \bibinfo{journal}{Phys. Fluids B} \textbf{\bibinfo{volume}{1}},
  \bibinfo{pages}{1331} (\bibinfo{year}{1989}).

\bibitem[{\citenamefont{Dorland and Hammett}(1993)}]{Dorland_Hammett_93}
\bibinfo{author}{\bibfnamefont{W.}~\bibnamefont{Dorland}} \bibnamefont{and}
  \bibinfo{author}{\bibfnamefont{G.~W.} \bibnamefont{Hammett}},
  \bibinfo{journal}{Phys. Fluids B} \textbf{\bibinfo{volume}{5}},
  \bibinfo{pages}{812} (\bibinfo{year}{1993}).

\bibitem[{\citenamefont{Hasegawa and Mima}(1977)}]{Hasegawa_Mima_77}
\bibinfo{author}{\bibfnamefont{A.}~\bibnamefont{Hasegawa}} \bibnamefont{and}
  \bibinfo{author}{\bibfnamefont{K.}~\bibnamefont{Mima}},
  \bibinfo{journal}{Phys. Rev. Lett.} \textbf{\bibinfo{volume}{39}},
  \bibinfo{pages}{205} (\bibinfo{year}{1977}).

\bibitem[{\citenamefont{Arakawa}(1966)}]{Arakawa_66}
\bibinfo{author}{\bibfnamefont{A.}~\bibnamefont{Arakawa}}, \bibinfo{journal}{J.
  Comput. Phys.} \textbf{\bibinfo{volume}{1}}, \bibinfo{pages}{119}
  (\bibinfo{year}{1966}).

\bibitem[{\citenamefont{Karniadakis et~al.}(1991)\citenamefont{Karniadakis,
  Israeli, and Orszag}}]{Karniadakis_Israeli_Orszag_91}
\bibinfo{author}{\bibfnamefont{G.~E.} \bibnamefont{Karniadakis}},
  \bibinfo{author}{\bibfnamefont{M.}~\bibnamefont{Israeli}}, \bibnamefont{and}
  \bibinfo{author}{\bibfnamefont{S.~A.} \bibnamefont{Orszag}},
  \bibinfo{journal}{J. Comput. Phys.} \textbf{\bibinfo{volume}{97}},
  \bibinfo{pages}{414} (\bibinfo{year}{1991}).

\bibitem[{\citenamefont{Numata et~al.}(2007)\citenamefont{Numata, Ball, and
  Dewar}}]{Numata_Ball_Dewar_07}
\bibinfo{author}{\bibfnamefont{R.}~\bibnamefont{Numata}},
  \bibinfo{author}{\bibfnamefont{R.}~\bibnamefont{Ball}}, \bibnamefont{and}
  \bibinfo{author}{\bibfnamefont{R.~L.} \bibnamefont{Dewar}}, in
  \emph{\bibinfo{booktitle}{Frontiers in Turbulence and Coherent Structures:
  Proceedings of the CSIRO/COSNet Workshop on Turbulence and Coherent
  Structures, Canberra, Australia, 10-13 January 2006}}, edited by
  \bibinfo{editor}{\bibfnamefont{J.~P.} \bibnamefont{Denier}} \bibnamefont{and}
  \bibinfo{editor}{\bibfnamefont{J.~S.} \bibnamefont{Frederiksen}}
  (\bibinfo{publisher}{World Scientific}, \bibinfo{address}{Singapore},
  \bibinfo{year}{2007}), vol.~\bibinfo{volume}{6} of
  \emph{\bibinfo{series}{World Scientific Lecture Notes in Complex Systems}},
  pp. \bibinfo{pages}{431--442}.

\bibitem[{\citenamefont{{Lord Rayleigh}}(1879)}]{Rayleigh_1880}
\bibinfo{author}{\bibnamefont{{Lord Rayleigh}}}, \bibinfo{journal}{Proc. London
  Math. Soc.} \textbf{\bibinfo{volume}{11}}, \bibinfo{pages}{57}
  (\bibinfo{year}{1879}).

\bibitem[{\citenamefont{Tollmien}(1935)}]{Tollmien_1935}
\bibinfo{author}{\bibfnamefont{W.}~\bibnamefont{Tollmien}},
  \bibinfo{journal}{Nachr. Ges. Wiss. G\"ottingen, Math.-Phys. Kl.}
  \textbf{\bibinfo{volume}{50}}, \bibinfo{pages}{79} (\bibinfo{year}{1935}).

\bibitem[{\citenamefont{Lin}(1945)}]{Lin_1945_II}
\bibinfo{author}{\bibfnamefont{C.~C.} \bibnamefont{Lin}},
  \bibinfo{journal}{Quart. Appl. Math.} \textbf{\bibinfo{volume}{3}},
  \bibinfo{pages}{218} (\bibinfo{year}{1945}).

\bibitem[{\citenamefont{Drazin and Reid}(1981)}]{Drazin_81}
\bibinfo{author}{\bibfnamefont{P.~G.} \bibnamefont{Drazin}} \bibnamefont{and}
  \bibinfo{author}{\bibfnamefont{W.~H.} \bibnamefont{Reid}},
  \emph{\bibinfo{title}{Hydrodynamic Stability}} (\bibinfo{publisher}{Cambridge
  University Press}, \bibinfo{address}{Cambridge}, \bibinfo{year}{1981}).

\bibitem[{\citenamefont{Kuo}(1949)}]{Kuo_1949}
\bibinfo{author}{\bibfnamefont{H.~L.} \bibnamefont{Kuo}}, \bibinfo{journal}{J.
  Met.} \textbf{\bibinfo{volume}{6}}, \bibinfo{pages}{105}
  (\bibinfo{year}{1949}).

\end{thebibliography}

\end{document}